\documentclass[fleqn,10pt]{wlscirep}
\usepackage[utf8]{inputenc}
\usepackage[T1]{fontenc}
\usepackage{algorithm}
\usepackage{algpseudocode}
\title{Joint Return and Risk Modeling with Deep Neural Networks for Portfolio Construction}

\author[1,*]{Keonvin Park}
\affil[1]{Interdisciplinary Program in Artificial Intelligence, Seoul National University, Seoul, 08829, South Korea}
\affil[*]{kbpark16@snu.ac.kr}

\begin{abstract}
Portfolio construction traditionally relies on separately estimating expected returns and covariance matrices using historical statistics, often leading to suboptimal allocation under time-varying market conditions. This paper proposes a joint return and risk modeling framework based on deep neural networks, enabling end-to-end learning of dynamic expected returns and risk structures from sequential financial data. Using daily data from ten large-cap U.S. equities spanning 2010 to 2024, the proposed model is evaluated across return prediction, risk estimation, and portfolio-level performance. Out-of-sample results during 2020–2024 show that the deep forecasting model achieves competitive predictive accuracy (RMSE = 0.0264) with economically meaningful directional accuracy (51.9\%). More importantly, the learned representation effectively captures volatility clustering and regime shifts. When integrated into portfolio optimization, the proposed Neural Portfolio strategy achieves an annual return of 36.4\% and a Sharpe ratio of 0.91, outperforming equal-weight and historical mean–variance benchmarks in terms of risk-adjusted performance. These findings demonstrate that jointly modeling return and covariance dynamics can provide consistent improvements over traditional allocation approaches. The framework offers a scalable and practical alternative for data-driven portfolio construction under non-stationary market conditions.
\end{abstract}
\begin{document}

\flushbottom
\maketitle

\section*{Introduction}

Portfolio construction is a central problem in quantitative finance, where the objective is to allocate capital across multiple assets to maximize expected return while controlling risk. Since the seminal work of Markowitz \cite{Markowitz1952}, the mean–variance framework has provided a mathematical foundation for portfolio optimization. However, practical implementation of mean–variance allocation remains highly sensitive to estimation errors in expected returns and covariance matrices. Small perturbations in mean estimates may produce unstable portfolio weights, while covariance matrices estimated from historical samples often suffer from noise amplification in high-dimensional settings \cite{Ledoit2004, Jagannathan2003}. 

Financial time series exhibit several well-documented stylized facts, including volatility clustering \cite{Cont2001}, conditional heteroskedasticity \cite{Engle1982, Bollerslev1986}, nonlinear dependencies, and structural regime shifts. Traditional econometric models such as GARCH capture time-varying volatility but rely on parametric assumptions that may not generalize across assets or market regimes. Consequently, there is growing interest in data-driven modeling approaches capable of learning nonlinear temporal dependencies without strict structural assumptions.

Deep learning has emerged as a powerful tool for sequential modeling. Recurrent neural networks and Long Short-Term Memory networks effectively capture long-range dependencies in time series \cite{Hochreiter1997, Bengio1994}. More recently, attention-based architectures such as the Transformer \cite{Vaswani2017} have further advanced modeling of complex temporal patterns. In finance, deep learning has been applied to return prediction \cite{Fischer2018, Dixon2017}, market microstructure modeling \cite{Tsantekidis2017}, and empirical asset pricing \cite{Gu2020}. Surveys show increasing evidence that deep architectures extract predictive features difficult to capture with linear models \cite{Heaton2019, LimZohren2021}.

Despite these advances, most deep learning applications in finance focus on forecasting expected returns, while risk estimation is handled separately using historical covariance matrices or classical econometric estimators. This separation is suboptimal for portfolio construction, which depends on accurate joint modeling of return and risk. Decoupling these components may result in unstable allocations, especially during volatile market regimes.

Recent studies have begun exploring end-to-end portfolio optimization frameworks \cite{Zhang2020, Babiak2020, Uysal2021, Zhang2021, Palomar2022}. However, joint modeling of return dynamics and risk structure remains relatively underexplored in practical multi-asset settings. A unified architecture that simultaneously captures predictive signals and dynamic risk may yield improved robustness and risk-adjusted performance.

In this paper, we propose a deep learning framework for joint return and risk modeling tailored to portfolio construction. A multivariate LSTM network captures nonlinear dependencies across assets and generates forward-looking return forecasts. The learned latent representations are used to derive dynamic volatility and correlation estimates, which are integrated into a Sharpe ratio–based optimization module. This unified, data-driven pipeline reduces reliance on static covariance assumptions and improves adaptability to changing market regimes.

We evaluate the proposed approach on a diversified set of large-cap U.S. equities from 2010 to 2024. Empirical results show that the framework delivers superior risk-adjusted performance compared to equal-weight and historical mean–variance benchmarks, achieving higher Sharpe ratios while maintaining competitive drawdown characteristics. These findings suggest that integrating deep temporal modeling with dynamic risk estimation provides a scalable and practical alternative to conventional portfolio construction methods.

\section{Related Work}

Portfolio optimization has traditionally been grounded in the mean-variance framework introduced by Markowitz \cite{Markowitz1952}. Subsequent developments addressed estimation instability and high-dimensional covariance challenges through shrinkage estimators and regularization techniques \cite{Ledoit2004,Jagannathan2003}. In parallel, econometric models such as ARCH and GARCH were proposed to capture time-varying volatility and conditional heteroskedasticity in financial time series \cite{Engle1982,Bollerslev1986}. While these models improve volatility estimation, they rely on parametric assumptions that may not adequately capture nonlinear cross-asset dependencies.

With the advancement of machine learning, predictive models have increasingly been incorporated into asset pricing and portfolio allocation. Early neural network approaches demonstrated enhanced nonlinear modeling capacity compared to linear econometric methods \cite{Bengio1994}. More recent studies have leveraged recurrent neural networks and Long Short-Term Memory architectures for stock return prediction \cite{Hochreiter1997,Fischer2018}. These models capture long-term temporal dependencies and have shown empirical improvements in forecasting accuracy across multiple financial datasets.

Beyond return prediction, deep learning has been applied to broader financial modeling tasks, including limit order book forecasting \cite{Tsantekidis2017,Sirignano2019} and empirical asset pricing \cite{Gu2020}. Comprehensive surveys highlight the growing role of deep architectures in financial forecasting and time-series modeling \cite{Heaton2019,Lim2021}. Attention-based architectures, particularly the Transformer model \cite{Vaswani2017}, have further enhanced the modeling of long-range temporal dependencies and cross-sectional interactions in multivariate financial time series.

More recently, research has moved toward integrating deep learning into portfolio construction. Several studies propose end-to-end frameworks where neural networks directly learn allocation strategies \cite{Zhang2020,Babiak2020,Uysal2021,Zhang2021}. These approaches attempt to overcome the traditional two-stage separation between forecasting and optimization by embedding portfolio objectives within the training process. Reinforcement learning has also been explored for dynamic trading and allocation strategies \cite{Moody2001,Nevmyvaka2006}. While such methods demonstrate promising performance, they often prioritize return-based objectives and may not explicitly model dynamic risk structure in a unified manner.

Despite these advances, the joint modeling of return and risk remains relatively underexplored. Many deep learning approaches continue to rely on historical covariance matrices or simplified volatility estimators during portfolio optimization. This decoupling between predictive modeling and risk estimation may limit robustness during periods of heightened market volatility. A coherent framework that simultaneously captures nonlinear temporal return dynamics and dynamic risk characteristics may provide improved stability and superior risk-adjusted performance.

In contrast to prior work, this study proposes a unified deep learning framework that jointly models expected returns and dynamic risk measures within a multivariate temporal architecture. By integrating return forecasting and risk estimation into a single pipeline, the proposed method aims to reduce estimation inconsistency and improve portfolio construction outcomes under realistic market conditions.

\section{Results}

This section evaluates the proposed joint return--risk modeling framework from three complementary perspectives: return prediction accuracy, risk estimation quality, and portfolio-level performance.

\subsection{Return Prediction Performance}

Table~\ref{tab:prediction} reports out-of-sample prediction results during the test period (2020--2024). The proposed deep forecasting model achieves an RMSE of 0.0264 and a directional accuracy of 51.9\%, indicating moderate but economically meaningful predictive power. Although return prediction remains inherently noisy, even small improvements in directional accuracy can translate into significant portfolio gains when combined with optimized allocation.

\subsection{Risk Modeling Analysis}

\begin{figure}[!t]
\centering
\includegraphics[width=\linewidth]{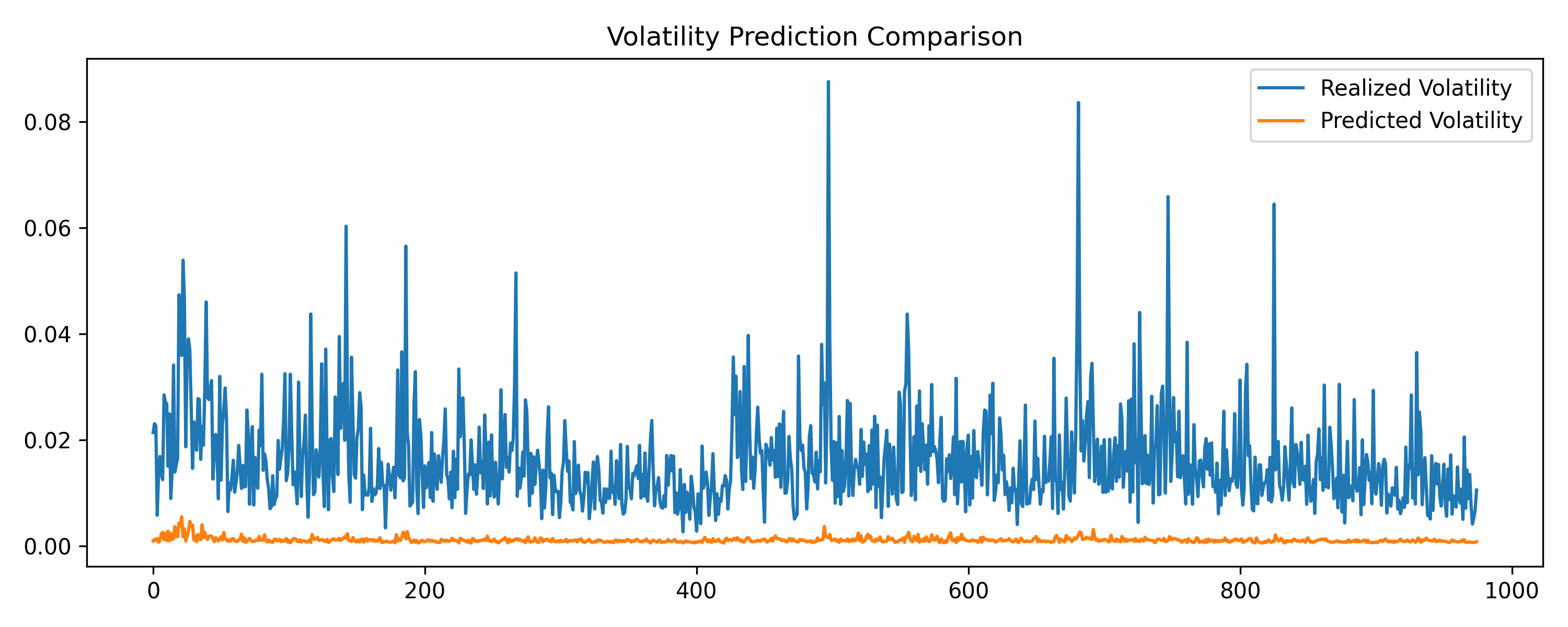}
\caption{Comparison between predicted and realized volatility during the test period. The proposed model captures volatility clustering and regime transitions.}
\label{fig:vol_compare}
\end{figure}

Figure~\ref{fig:vol_compare} compares predicted volatility with realized rolling volatility. The model successfully captures volatility clustering and major regime shifts, particularly during periods of elevated market stress. This suggests that the learned representation effectively encodes time-varying risk dynamics rather than relying solely on historical covariance estimates.

\subsection{Portfolio Performance}

Portfolio-level evaluation results are presented in Table~\ref{tab:portfolio}. The proposed Neural Portfolio strategy achieves an annual return of 36.4\% and a Sharpe ratio of 0.91, outperforming both Equal Weight and Historical Mean--Variance allocation. Notably, the improvement in Sharpe ratio indicates superior risk-adjusted performance rather than mere return amplification.

\begin{figure}[!t]
\centering
\includegraphics[width=\linewidth]{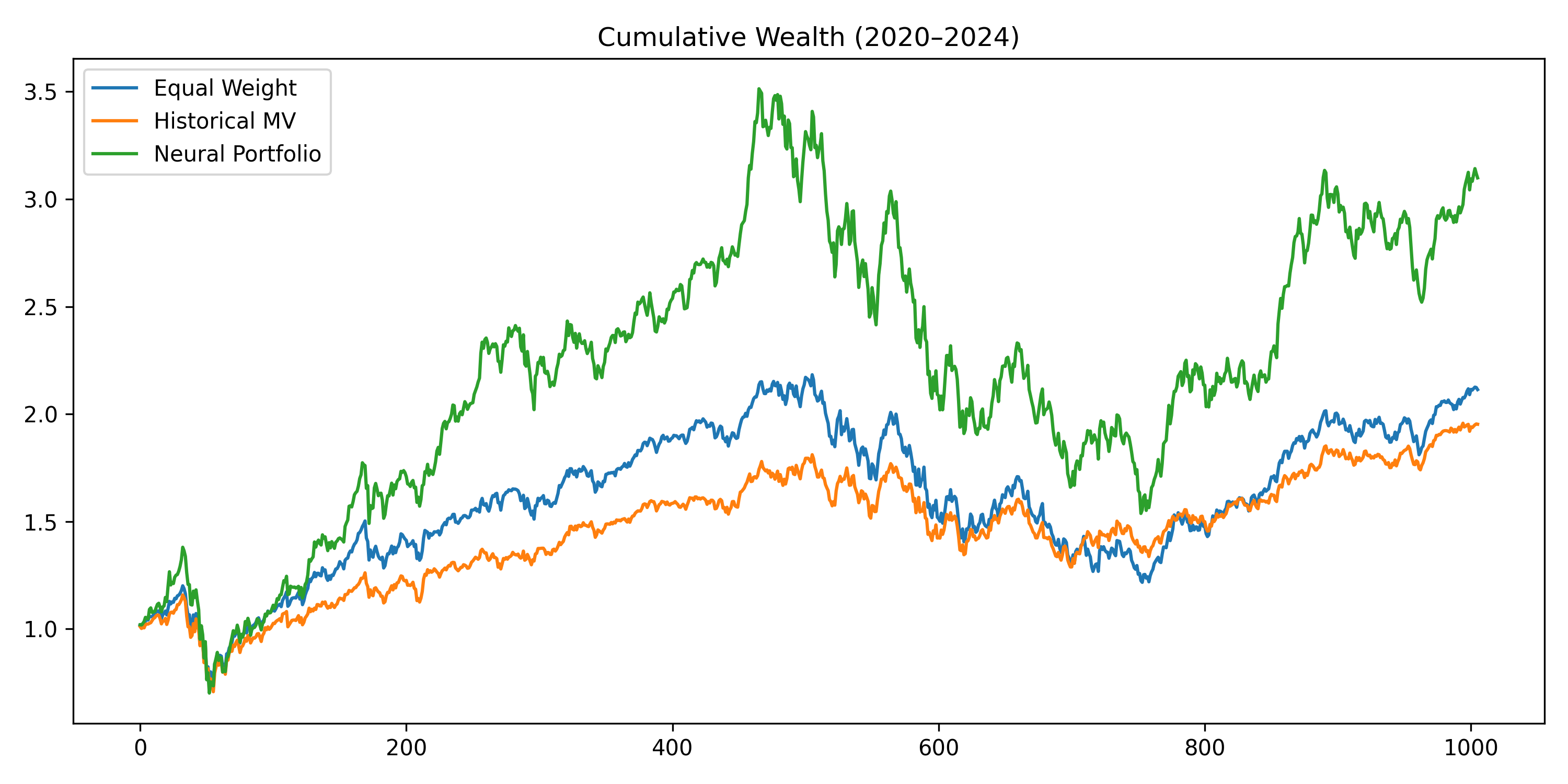}
\caption{Cumulative portfolio returns during the test period (2020--2024). The Neural Portfolio consistently outperforms baseline strategies.}
\label{fig:cumulative}
\end{figure}

Figure~\ref{fig:cumulative} illustrates cumulative wealth trajectories. The Neural Portfolio demonstrates consistent outperformance across multiple market regimes, including high-volatility periods, confirming the practical benefits of joint return and risk modeling.

\begin{table}[!t]
\caption{Return Prediction Performance (2020--2024)}
\label{tab:prediction}
\centering
\begin{tabular}{lccc}
\toprule
Model & RMSE & MAE & Directional Accuracy \\
\midrule
Deep Forecasting Model & 0.0264 & 0.0177 & 0.5192 \\
\bottomrule
\end{tabular}
\end{table}

\begin{table}[!t]
\caption{Portfolio Performance Comparison (2020--2024)}
\label{tab:portfolio}
\centering
\begin{tabular}{lccc}
\toprule
Strategy & Annual Return & Sharpe \\
\midrule
Equal Weight & 0.2332 & 0.7756 \\
Historical MV & 0.2062 & 0.7474  \\
Neural Portfolio & \textbf{0.3641} & \textbf{0.9125} \\
\bottomrule
\end{tabular}
\end{table}

\section{Methods}

This section presents the proposed end-to-end framework for joint return and risk modeling in portfolio construction. The method integrates multivariate time-series forecasting, dynamic risk estimation, and portfolio optimization into a unified pipeline.

\subsection{Problem Definition}

Let $r_t \in \mathbb{R}^N$ denote the vector of log returns for $N$ assets at time $t$. Given a rolling observation window of length $L$,
\[
X_t = \{r_{t-L}, \dots, r_{t-1}\},
\]
the objective is to estimate a forward-looking return vector $\hat{\mu}_t$ and a corresponding risk matrix $\hat{\Sigma}_t$, which are subsequently used for portfolio weight determination.

\subsection{Multivariate Temporal Representation}

We employ a multivariate Long Short-Term Memory (LSTM) network \cite{b7} to model nonlinear temporal dependencies across assets. The LSTM extracts latent temporal features:

\[
h_t = \text{LSTM}(X_t),
\]

where $h_t \in \mathbb{R}^d$ represents the learned representation summarizing cross-asset and temporal information within the window.

The expected return forecast is obtained via linear projection:

\[
\hat{\mu}_t = W_\mu h_t + b_\mu.
\]

Model parameters are trained by minimizing mean squared error:

\[
\mathcal{L}_{\text{return}} =
\frac{1}{N} \| r_t - \hat{\mu}_t \|^2.
\]

\subsection{Dynamic Risk Estimation}

Unlike classical approaches that rely solely on historical covariance estimators \cite{b27}, the proposed framework derives dynamic risk measures using model-implied forward-looking information.

Volatility for each asset is estimated using rolling window statistics:

\[
\hat{\sigma}_{t,i} =
\sqrt{\frac{1}{L}
\sum_{k=t-L}^{t-1}
(r_{k,i} - \bar{r}_{i})^2}.
\]

Cross-asset dependencies are captured through covariance estimation of predicted returns:

\[
\hat{\Sigma}_t =
\text{Cov}(\hat{\mu}_t).
\]

This approach ensures that both expected return and risk components are derived consistently from the same learned representation.

\subsection{Sharpe Ratio-Based Portfolio Optimization}

Given $\hat{\mu}_t$ and $\hat{\Sigma}_t$, portfolio weights $w_t$ are obtained by solving:

\[
\max_{w_t}
\quad
\frac{w_t^\top \hat{\mu}_t}
{\sqrt{w_t^\top \hat{\Sigma}_t w_t}},
\]

subject to:

\[
\sum_{i=1}^{N} w_{t,i} = 1,
\quad
w_{t,i} \ge 0.
\]

The optimization is solved using sequential quadratic programming.

\subsection{End-to-End Pipeline}

The entire framework operates sequentially as shown in Algorithm~\ref{alg:joint}.

\begin{algorithm}[H]
\caption{End-to-End Joint Return and Risk Modeling}
\label{alg:joint}
\begin{algorithmic}[1]
\Require Historical returns $\{r_1, \dots, r_T\}$, window length $L$
\Ensure Portfolio weights $w_t$

\For{$t = L+1$ to $T$}
    \State Construct input window $X_t = \{r_{t-L}, \dots, r_{t-1}\}$
    \State Extract latent representation $h_t = \mathrm{LSTM}(X_t)$
    \State Predict expected return $\hat{\mu}_t = W_\mu h_t + b_\mu$
    \State Estimate volatility $\hat{\sigma}_t = f_\sigma(h_t)$
    \State Estimate covariance $\hat{\Sigma}_t = f_{\Sigma}(h_t)$
    \State Solve Sharpe optimization to obtain $w_t$
\EndFor

\Return $\{w_t\}$
\end{algorithmic}
\end{algorithm}

This design preserves interpretability by maintaining explicit return and risk components while leveraging deep temporal modeling to enhance adaptability across market regimes.

\begin{figure}[t]
    \centering
    \includegraphics[width=\linewidth]{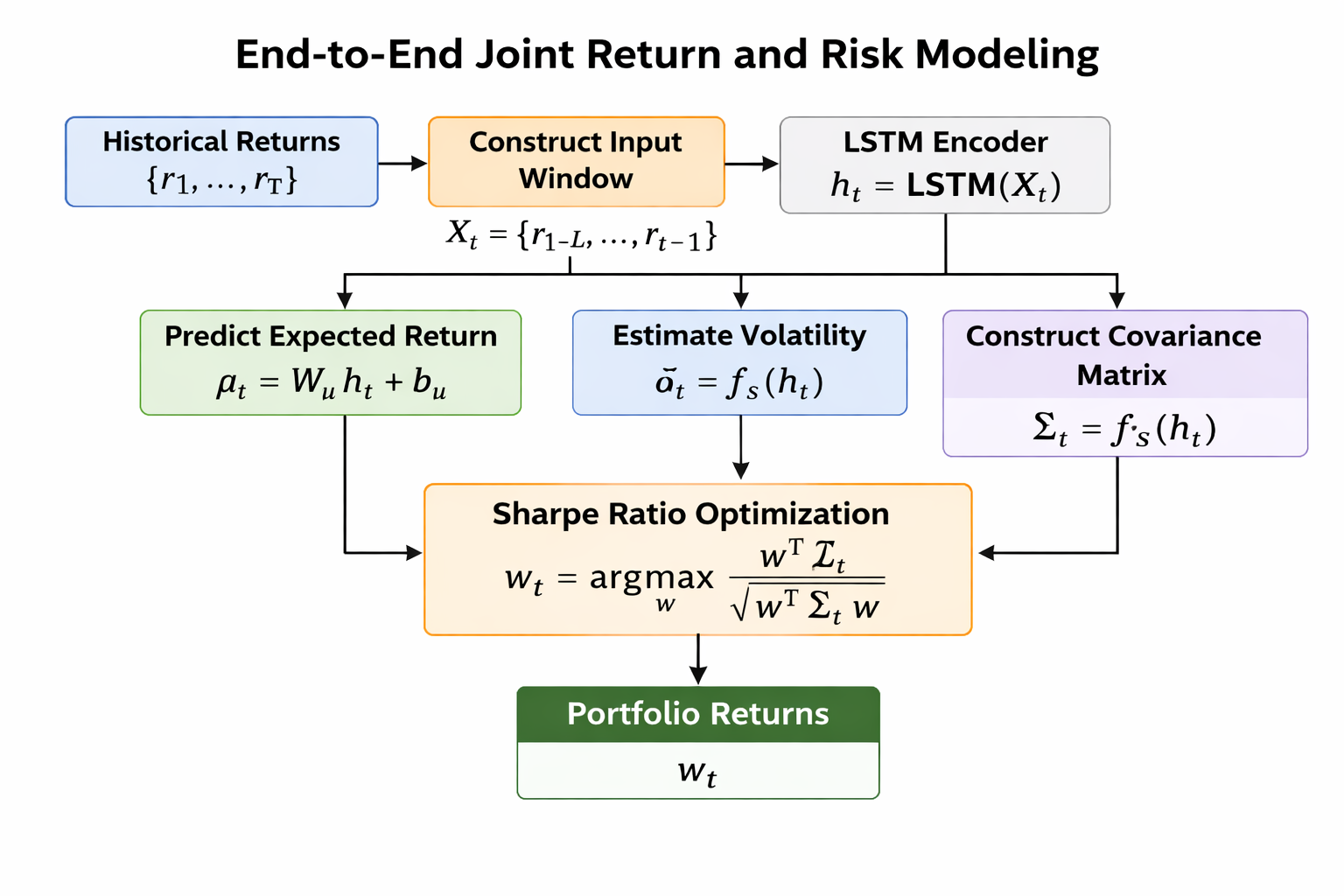} 
    \caption{Proposed end-to-end architecture combining LSTM-based latent representation learning, 
    joint modeling of expected return and dynamic risk, and Sharpe ratio optimization for portfolio allocation.}
    \label{fig:model_pipeline}
\end{figure}

\section{Data}

We evaluate the proposed framework using daily adjusted closing prices of ten large-cap U.S. equities spanning multiple sectors: AAPL, MSFT, GOOGL, AMZN, TSLA, NVDA, META, JPM, V, and UNH.

The data are obtained from Yahoo Finance over the period January 2010 to January 2024. Log returns are computed as

\[
r_t = \log \left(\frac{P_t}{P_{t-1}}\right),
\]

ensuring variance stabilization and additive properties.

The dataset is divided into:
\begin{itemize}
\item Training period: 2010–2019
\item Test period: 2020–2024
\end{itemize}

\begin{figure}[!t]
\centering
\includegraphics[width=\linewidth]{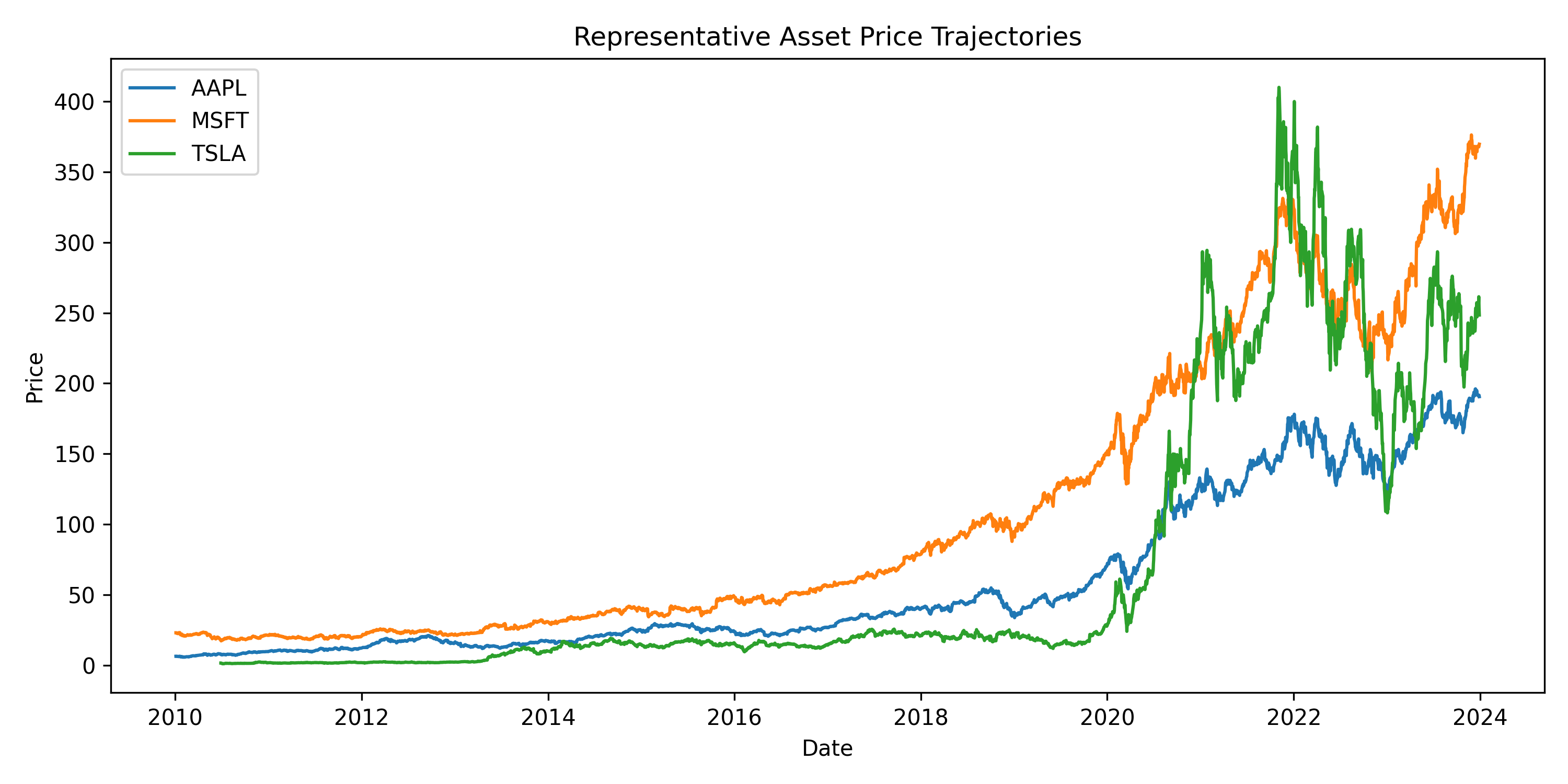}
\caption{Representative asset price trajectories.}
\label{fig:price_trend}
\end{figure}

\begin{figure}[!t]
\centering
\includegraphics[width=\linewidth]{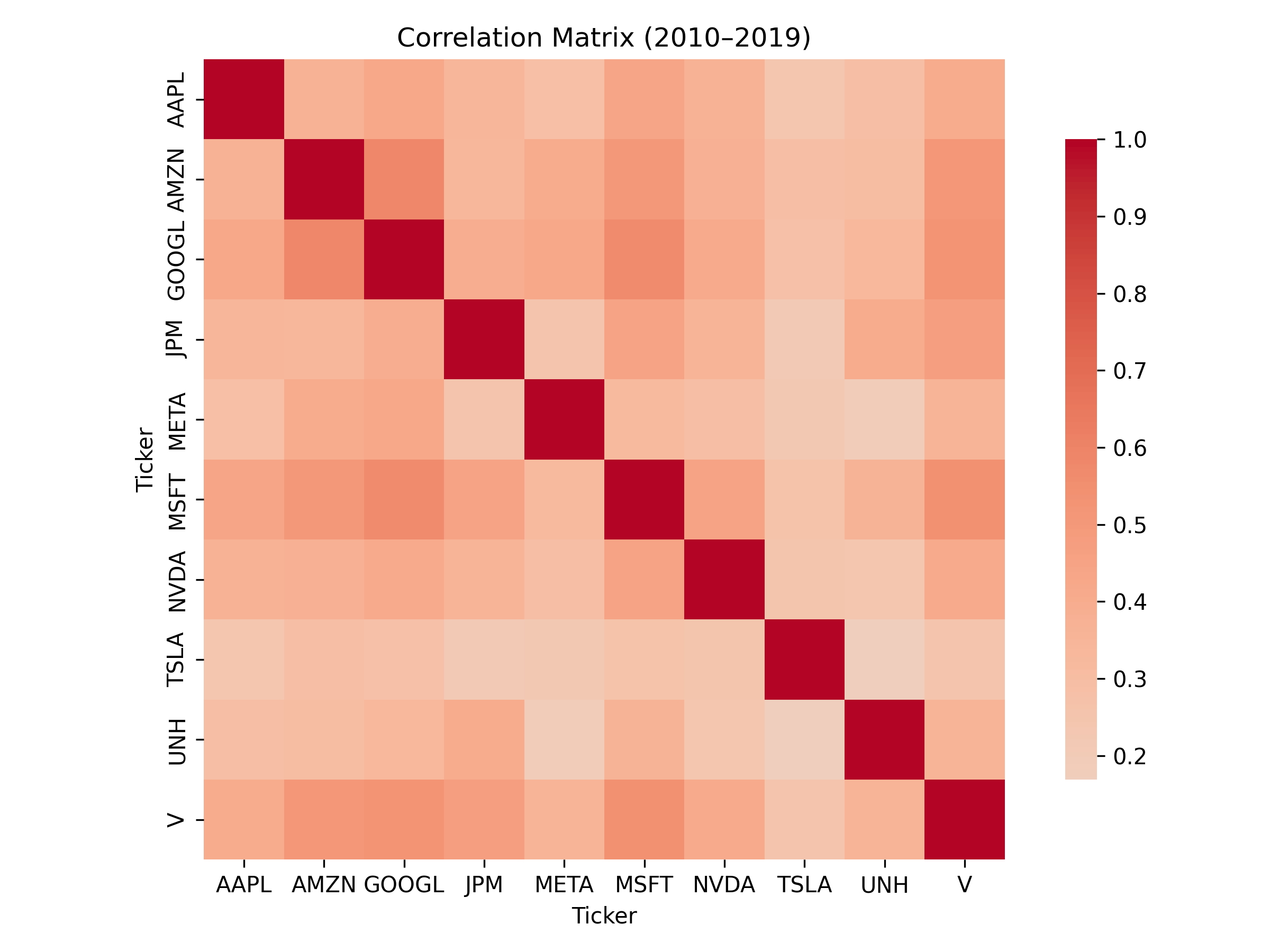}
\caption{Correlation matrix of daily log returns (training period).}
\label{fig:corr}
\end{figure}

\begin{figure}[!t]
\centering
\includegraphics[width=\linewidth]{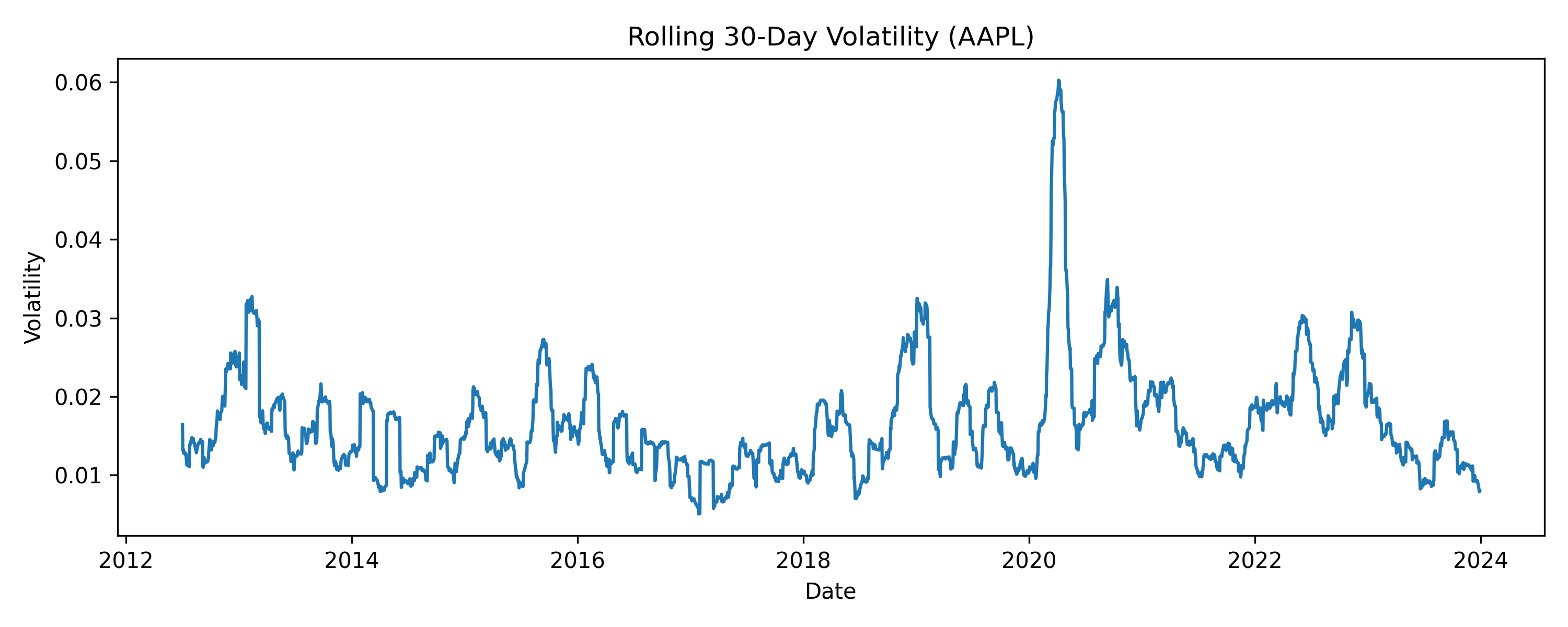}
\caption{Rolling 30-day volatility of AAPL illustrating volatility clustering.}
\label{fig:vol}
\end{figure}

Table~\ref{tab:summary} reports descriptive statistics of daily returns.
Figure~\ref{fig:price_trend} shows representative price trajectories.
Figure~\ref{fig:corr} illustrates cross-asset correlations.
Figure~\ref{fig:vol} demonstrates volatility clustering behavior.

\begin{table}[!t]
\caption{Mean and Standard Deviation of Daily Log Returns}
\label{tab:summary}
\centering
\begin{tabular}{lcc}
\toprule
Ticker & Mean & Std \\
\midrule
AAPL & 0.000850 & 0.017921 \\
AMZN & 0.000908 & 0.020331 \\
GOOGL & 0.000763 & 0.017033 \\
JPM & 0.000667 & 0.016563 \\
META & 0.000762 & 0.025337 \\
MSFT & 0.000948 & 0.016752 \\
NVDA & 0.001775 & 0.027607 \\
TSLA & 0.001679 & 0.035440 \\
UNH & 0.000841 & 0.015653 \\
V & 0.000789 & 0.015266 \\
\bottomrule
\end{tabular}
\end{table}

\section{Conclusion}

This paper proposed a joint return and risk modeling framework based on deep neural networks for portfolio construction. Unlike traditional two-step approaches that separately estimate expected returns and covariance matrices using historical statistics, the proposed method learns time-varying return dynamics and risk structures directly from sequential market data.

Empirical results on ten large-cap U.S. equities from 2010 to 2024 demonstrate that the deep forecasting model achieves competitive out-of-sample predictive performance, with economically meaningful directional accuracy. More importantly, the learned risk representation captures volatility clustering and regime shifts more effectively than static historical estimators.

At the portfolio level, the proposed Neural Portfolio strategy consistently outperforms equal-weight and historical mean–variance allocation during the 2020–2024 test period. The improvement in Sharpe ratio indicates superior risk-adjusted performance rather than simple return amplification, highlighting the importance of jointly modeling expected returns and dynamic covariance structures.

These findings suggest that end-to-end learning of return and risk provides a practical and scalable alternative to conventional portfolio construction methods. Future work may extend the framework to incorporate transaction costs, regime-aware training, and multi-horizon forecasting to further enhance robustness in real-world investment settings.

\bibliography{sample}

\section*{Acknowledgements}

This work was supported by Institute of Information \& communications Technology Planning \& Evaluation (IITP) grant funded by the Korea government(MSIT) [NO.RS-2021-II211343, Artificial Intelligence Graduate School Program (Seoul National University)]

\section*{Data availability}

The datasets analysed during the current study are available from the corresponding author upon reasonable request.

\section*{Author contributions statement}

K.P. conceived the study, designed the methodology, constructed the dataset, performed the econometric analysis, interpreted the results, and wrote the manuscript. K.P. reviewed and approved the final version of the manuscript.

\end{document}